# Software Defect Prediction by Online Learning Considering Defect Overlooking

Yuta Yamasaki
*Kindai University*
Higashi-osaka, Japan
2133340459f@kindai.ac.jp

Nikolay Fedorov
*Dubna State University*
Dubna, Russia

Masateru Tsunoda
Kindai University
Higashi-osaka, Japan
tsunoda@info.kindai.ac.jp

Akito Monden
Okayama University
Okayama, Japan
monden@okayama-u.ac.jp

Amjed Tahir
Massey University
Palmerston North, New Zealand
a.tahir@massey.ac.nz

Kwabena Ebo Bennin
*Wageningen UR*
Wageningen, Netherlands
kwabena.bennin@wur.nl

Koji Toda
*Fukuoka Institute of Technology*
Fukuoka, Japan
toda@fit.ac.jp

Keitaro Nakasai
*OMU College of Technology*
Osaka, Japan
nakasai@omu.ac.jp

*Abstract*—Building defect prediction models based on online learning can enhance prediction accuracy. It continuously rebuilds a new prediction model when adding a new data point. However, predicting a module as "non-defective" (i.e., negative prediction) can result in fewer test cases for such modules. Therefore, defects can be overlooked during testing, even when the module is defective. The erroneous test results are used as learning data by online learning, which could negatively affect prediction accuracy. In our experiment, we demonstrate this negative influence on prediction accuracy.

## I. Introduction

Software testing is one of the key activities to find defects. However, due to resource availability and software development duration, testing can be limited to only a few modules [6]. Defect prediction is one of the major approaches to suppressing remaining defects. When a module is regarded as defective by the prediction model, it is tested thoroughly (i.e., more effort is spent on testing it). In contrast, when regarded as non-defective, it can be tested much more lightly [2]. When the accuracy of the prediction model is high, both low testing costs and high software quality can be achieved.

Learning data based on the previous versions' history is often used to build a defect prediction model. For instance, during the development of version 1.0, data such as the number of found defects and the complexity of the modules are recorded. Next, a defect prediction model for the next version (e.g., 1.1) is built using this data. Lastly, during the development of version 1.1 (i.e., on test data), defects of each module are predicted using the prediction model built in the previous stage. The procedure is called cross-version defect prediction (CVDP).

However, the accuracy of CVDP is often low. This is because when the version is different between learning and test data, the independent variables of the prediction model are often different. This is regarded as an external validity issue of defect prediction [1]. To address the problem, online learning approaches have been proposed recently [4]. When a new data point is added, online learning adds it to the learning dataset and rebuilds a new prediction model. Using this approach, software testing results are collected and utilized to enhance prediction accuracy during development.

Fig. 1 illustrates an example of defect prediction by online learning. Each module is tested sequentially from the top to the bottom. After module t9 is tested (i.e., before t5), a prediction model M1 is built. The learning dataset includes modules t1 and t9, where an independent variable is the lines of code (LOC), and a dependent variable is the test result. M1 predicts the test result of t5. After t5 is tested, model M2 is built based on t1, t9, and t5 data. M2 predicts the test result of t7.

## II. Defect Overlooking

When a defect prediction model predicts a negative result (i.e., "non-defective"), developers will typically write fewer test cases for those modules [2] to efficiently allocate testing resources [3][6]. As a result, the test overlooks defects, and the module might be regarded as "non-defective" in most cases, even if the module is defective. We call this case a *defect overlooking by negative prediction*. This means defects are overlooked due to fewer test cases based on negative prediction.

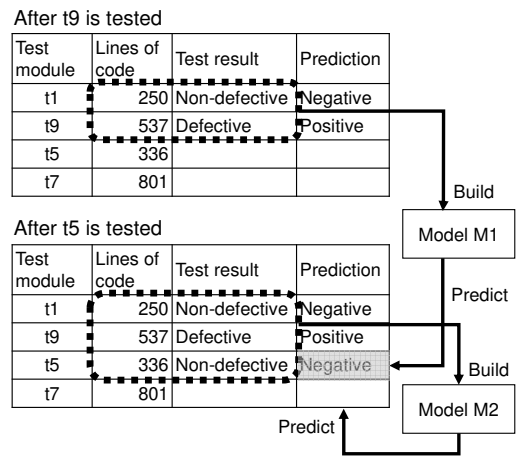

Fig. 1. **Example of defect prediction by online learning**

| Test module | Lines of code | Test result | Prediction | Actual result After testing |
|---|---|---|---|---|
| t1 | 250 | Non-defective | Negative | **Defective** |
| t9 | 537 | Defective | Positive | Defective |
| t5 | 336 | Non-defective | Negative | **Defective** |
| t7 | 801 | Non-defective | Negative | **Defective** |

Fig. 2. **Example of defect overlookng**

Such overlooking of defects could negatively affect the accuracy of the prediction models. In Figure 2, the column "test result" considers only defects during testing, while "actual result after testing" also considers defects after testing was done (e.g., when the software is released). In the example, we assume defects are always overlooked when the prediction is negative due to fewer test cases. That is, when the "Prediction" column is "Negative" in Figure 2, the "Test result" column is "Non-defective" with 100% probability.

Module t1 and t5 are regarded as non-defective based on the test outcomes, and they are used as learning data for model M2. However, based on the actual result, the learning data must be corrected and set as defective. As a result, the accuracy of model M2 becomes low, and the model predicts module t7 as "non-defective" erroneously.

The defects' overlooking issue was not considered in previous defect prediction by online learning studies. Without considering such overlooking, the accuracy of these prediction models might be evaluated incorrectly. However, in our previous work, we pointed out the influence of the overlooking by negative prediction; the study evaluated the influence on online model selection (i.e., models are not rebuilt during software testing). Therefore, it is still unclear what influence it may have on the accuracy of rebuilt models with online learning.

## III. EXPERIMENT

**Settings**: In the experiment, we assume that defect overlooking occurs with $n\%$ probability when a defect prediction model predicts "non-defective." We set $n$ as 0, 80 and 100. For instance, in Fig. 2, when $n$ is 80, test result of t1, t5, and t7 becomes "non-defective" at 80% probability. We set $n$ as 0 for the baseline, which does not consider the overlooking. In addition, when defects of the first module (e.g., t1 on Fig. 1) are predicted, there is no learning dataset. Therefore, we fixed the prediction as "defective" on the module.

When evaluating the models, we randomly sorted the order of modules 10 times and calculated the average of the evaluation criteria acquired from the 10 repetitions. This is because the prediction accuracy of online learning affects the order of module testing.

**Dataset**: We selected 3 datasets published on PROMISE and D'Ambros et al. [1] repositories to perform our cross-version defect prediction. Each dataset includes 20 independent variables, which include product metrics such as CK metrics. Table 1 shows details of the datasets used in the experiment.

**Prediction model**: To predict defective modules, we used logistic regression, a widely used method in defect prediction. As a feature selection method, we applied correlation-based feature selection, which is shown to be effective when used with logistic regression.

TABLE I.  DATASETS USED IN THE EXPERIMENT

| Software | Learning dataset | | | Test dataset | | |
|---|---|---|---|---|---|---|
| | Ver. | # of Modules (%) | | Ver. | # of Modules (%) | |
| | | All | Defective | | All | Defective |
| ant | 1.6 | 351 | 92 (26.2) | 1.7 | 745 | 166 (22.3) |
| prop | 5 | 8516 | 1299 (15.3) | 6 | 660 | 66 (10.0) |
| synapse | 1.1 | 222 | 60 (27.0) | 1.2 | 256 | 86 (33.6) |

TABLE II.  RELATIONSHIP BETWEEN PREDICTION ACCURACY AND PROBABILITY OF OVERLOOKING

| Software | Probability of overlooking (%) | AUC | F1 score |
|---|---|---|---|
| ant | 0 (baseline) | 0.74 | 0.56 |
| | 80 | 0.56 | 0.28 |
| | 100 | 0.55 | 0.46 |
| prop | 0 (baseline) | 0.63 | 0.26 |
| | 80 | 0.53 | 0.16 |
| | 100 | 0.49 | 0.11 |
| synapse | 0 (baseline) | 0.70 | 0.61 |
| | 80 | 0.55 | 0.30 |
| | 100 | 0.51 | 0.26 |

**Evaluation criteria**: We used AUC and F1 score to evaluate the performance of each prediction model. Prediction values are real numbers in most models. Hence, the prediction values were converted into binary values (i.e., defective or not defective) based on a cutoff value. We set the cutoff value as the closest point to the top left corner of the ROC curve on a training project. The converted values were also used to calculate the F1 score.

**Result**: The average prediction accuracy of the models is shown in Table 2. In the table, when the probability of overlooking is 100, the accuracy is very low compared with the baseline, which does not consider the overlooking. Even when the probability is 80, the accuracy is still lower than the baseline. The result suggests that to evaluate the accuracy of online learning, we have to consider the influence of overlooking.

ACKNOWLEDGMENT

This research is partially supported by the JSPS [Grants-in-Aid for Scientific Research (C) (No.21K11840).

REFERENCES

[1] M. D'Ambros, M., Lanza, and R. Robbes, "Evaluating defect prediction approaches: a benchmark and an extensive comparison," Empirical Software Engineering, vol.17, no.4-5, pp.531-577, 2012.

[2] S. Mahfuz, Software Quality Assurance - Integrating Testing, Security, and Audit, CRC Press, 2016.

[3] M. Shepperd, D. Bowes, and T. Hall, "Researcher Bias: The Use of Machine Learning in Software Defect Prediction," IEEE Transactions on Software Engineering, vol.40, no.6, pp.603-616, 2014.

[4] S. Tabassum, L. Minku, D. Feng, G. Cabral and L. Song, "An Investigation of Cross-Project Learning in Online Just-In-Time Software Defect Prediction," Proc. of International Conference on Software Engineering (ICSE), pp.554-565, 2020.

[5] M. Tsunoda, A. Monden, K. Toda, A. Tahir, K. Bennin, K. Nakasai, M. Nagura, and K. Matsumoto, "Using Bandit Algorithms for Selecting Feature Reduction Techniques in Software Defect Prediction," Proc. of Mining Software Repositories Conference (MSR), pp.670-681, 2022.

[6] T. Zimmermann and N. Nagappan, "Predicting defects using network analysis on dependency graphs," Proc. of International Conference on Software Engineering (ICSE), pp.531-540, 2008.